\documentclass{article}
\usepackage{spconf,amsmath,epsfig,amssymb,xcolor, float}

\newcommand{\proba}{\text{p}}
\DeclareMathOperator*{\argmin}{argmin}

\newcommand{\speckle}{{u}}
\newcommand{\reflectivity}{{v}}
\newcommand{\intensity}{{w}}
\newcommand{\logreflectivity}{{x}}
\newcommand{\logintensity}{{y}}
\newcommand{\logspeckle}{{z}}
\newcommand{\ratio}{{\tau}}
\newcommand{\superimage}{{s}}

\newcommand{\Emanuele}[1]{\textcolor{black}{#1}}

\title{EXPLOITING MULTI-TEMPORAL INFORMATION FOR IMPROVED SPECKLE REDUCTION OF SENTINEL-1 SAR IMAGES BY DEEP LEARNING}
\name{Emanuele Dalsasso$^\dagger$, Inès Meraoumia$^\dagger$\thanks{This work has been partly funded by Futur and Rupture Program of Institut Mines-Télécom.}, Loïc Denis$^\ddagger$, Florence Tupin$^\dagger$}
\address{$^1$LTCI, Télécom Paris, Institut Polytechnique de Paris, Palaiseau, France \\ $^\ddagger$Univ Lyon, UJM-Saint-Etienne, CNRS,
Institut d Optique Graduate School,\\ Laboratoire Hubert Curien UMR
5516, F-42023, SAINT-ETIENNE, France\\ }
\let\OLDthebibliography\thebibliography
\renewcommand\thebibliography[1]{
  \OLDthebibliography{#1}
  \setlength{\parskip}{2pt}
  \setlength{\itemsep}{0pt plus 0.3ex}
}

\begin{document}
%
\maketitle
\begin{abstract}
Deep learning approaches show unprecedented results for speckle reduction in SAR amplitude images. 
The wide availability of multi-temporal stacks of SAR images can improve even further the quality of denoising. In this paper, we propose 
a flexible yet efficient way to integrate temporal information into a deep neural network for speckle suppression. 
Archives provide access to long time-series of SAR images, from which multi-temporal averages can be computed with virtually no remaining speckle fluctuations.
The proposed method combines this multi-temporal average and the image at a given date in the form of a ratio image and uses a state-of-the-art neural network to remove the speckle in this ratio image. This simple strategy is shown to offer a noticeable improvement compared to filtering the original image without knowledge of the multi-temporal average.
\end{abstract}
\section{Introduction}
\label{sec:intro}
Remote sensing technologies are widely used for Earth observation, with a stream of images that are continuously provided by satellites carrying instruments with imaging capabilities. Among them, Synthetic Aperture Radar (SAR) is an active-imaging technology particularly attractive due to its capability to operate by day or night, in (almost) any weather condition, and to the complementary information it provides in addition to optical imaging. Interpreting SAR images is however not trivial. Indeed, they are affected by 
strong fluctuations
inherent to coherent systems: the so-called \textit{speckle} phenomenon. 



The speckle has been modeled by Goodman as a multiplicative noise \cite{Good-76}, whose variance is signal-dependent. 
Speckle reduction techniques have been intensively studied. 
Most recent approaches rely on cutting-edge machine learning techniques based on deep neural networks. These networks have the ability to learn implicit patterns of the set of images 
used during training and then to generalize to similar data. Supervised techniques require a large amount of noise-free and noisy image pairs. However, for SAR images, providing ground-truth images is difficult.
To circumvent this problem, MuLoG \cite{Dele-17} employs pre-trained Gaussian denoisers trained on natural images inside a framework that alternates non-linear steps, accounting for the Fisher-Tippett distribution of log-transformed data, and Gaussian denoising steps. Alternatively, noisy images resembling SAR images can be generated by simulating speckle noise on optical images, which then serve as ground-truth \cite{wang2017sar,wang2017generative,zhang2018learning}. As the discrepancy between the optical domain and the SAR domain is a source of artifacts, multi-temporal SAR series can instead be averaged to produce an image whose content  better matches that of an actual SAR acquisition (e.g. textures, edges, bright points). This image can then serve as reference with respect to another SAR image acquired on the same area \cite{chierchia2017sar,cozzolino2020nonlocal}. 
Synthetic speckle noise can also be added to the average 
image \cite{lattari2019deep} or to a denoised version of the average image
\cite{dalsasso2020sar}. The disadvantages stemming from the lack of an actual ground-truth (and above all, the difficulty to accurately model speckle correlations) motivate the interest of the community towards self-supervised approaches \cite{molini2020towards,molini2020speckle2void,ma2020sar,dalsasso2020sar2sar}. 


Since the launch of the Sentinel-1 satellite mission in the context of the Copernicus program of the European Space Agency, long time-series of SAR images are nowadays freely available online.
\Emanuele{Incorporating multi-temporal information with deep learning can potentially significantly improve denoising processes. Notably, stable structures in observed areas such as buildings or roads are clearly visible in multi-temporal averages, even when the contrast with the surrounding area is weak. RABASAR \cite{rabasar} is a framework designed to extend single-image speckle reduction techniques by including a high-quality multi-temporal average image. The multi-temporal mean is referred to as a "super-image", owing to its high signal-to-noise ratio, and represents a summary of the multi-temporal SAR stack. RABASAR combines a speckled image with the super-image by forming the ratio between the two images. This ratio image has a reduced informational content (only speckle fluctuations and changes with respect to the super-image are present) and is thus easier to denoise.
Most of the single-image despeckling methods are designed for spatially uncorrelated speckle and require a spatial resampling (typically, a down-sampling by a factor 2) when applied to actual SAR data.
Utilizing SAR2SAR \cite{dalsasso2020sar2sar}, a self-supervised deep learning algorithm, avoids sub-sampling altogether which preserves the spatial resolution of the restored images. Indeed, SAR2SAR is trained directly on Sentinel-1 SAR images, integrating knowledge of the spatial correlation of speckle. The main idea of this paper is to combine multi-temporal information with the SAR2SAR deep learning algorithm to improve speckle reduction in SAR imaging.}

\section{The SAR2SAR approach}
\label{sec:sar2sar}
Under the hypothesis of a sufficiently high number of independent and identically distributed scatterers, Goodman's fully-developed model \cite{Good-76} relates the measured intensity $\intensity$, the reflectivity $\reflectivity$, and the speckle $\speckle$ in a multiplicative model: $\intensity = \reflectivity \times \speckle $.
The speckle $\speckle$ is statistically described as a random variable following a gamma distribution:
\begin{equation}
    \proba(\speckle) = \frac{L^L}{\Gamma(L)}\speckle^{L-1}\exp{(-L\speckle)}\,,
\end{equation}
depending on the number of looks $L\geq 1$. It follows that the intensity $\intensity$ has a signal-dependent expectation $\mathbb{E}[\intensity]=\reflectivity$ and variance $\text{Var}[\intensity]=\reflectivity^2/L$. 
When training a neural network for speckle reduction, it is often beneficial to apply a logarithmic transform to input SAR intensities ($\logintensity=\log \intensity$). Indeed, not only does it allow to compress the high-dynamic range of values, but it also stabilizes the variance. The log-speckle $\logspeckle$ then has an additive behaviour
    $\logintensity = \logreflectivity + \logspeckle$ 
and is distributed according to a Fisher-Tippett distribution:
\begin{equation}
    \proba(\logspeckle) = \frac{L^L}{\Gamma(L)}\exp{(L \logspeckle)} \exp{(Le^\logspeckle)}\,.
\end{equation}
After a large dataset composed by pairs of ground-truth ($\logreflectivity_i$) and noisy ($\logintensity_i$) images is built, a Convolutional Neural Network (CNN) can be trained for speckle reduction to learn a parametric mapping $f_\theta$ that minimizes the empirical risk according to a loss function $\mathcal L$:
\begin{equation}
    \argmin_\theta \sum_i \mathcal L\left\{f_\theta(\logintensity_i),\logreflectivity_i\right\}\,.
\end{equation}
It is shown in \cite{dalsasso2020sar2sar} that the negative log-likelihood defines an efficient loss, given, for the Fisher-Tippett distribution, by:
\begin{equation}\label{eq:supervised_loss}
    \mathcal L^{lik}\left\{f_\theta(\logintensity_i),\logreflectivity_i\right\} = f_\theta(\logintensity_i)-\logreflectivity_i+\exp{\left(\logreflectivity_i-f_\theta(\logintensity_i)\right)}\,.
\end{equation}
In practice, given the impossibility to access to the ground-truth reflectivity, multi-temporal stacks of co-registered SAR images can be used instead. Given a set $\{\logintensity_1,\logintensity_2,\dots\}$ of co-registered SAR images, the network parameters $\theta$ can be learned by:
\begin{equation}
    \argmin_\theta 
    \; \sum_i \sum_{j\neq i}\mathcal L^{lik}\left\{f_\theta(\logintensity_i),\logintensity_j\right\}\,,
\end{equation}
provided that temporal changes between images $y_i$ and $y_j$ are adequately compensated. 
This unsupervised training is at the core of the SAR2SAR algorithm \cite{dalsasso2020sar2sar}, where multi-temporal stacks of Sentinel-1 SAR images are used to train an unsupervised model for speckle suppression from 1-look images. 
The advantage of learning directly from real SAR images is that the network can model the spatial correlations of speckle. Thus, the algorithm can be directly applied to SAR images without any pre-processing step (which preserves the spatial resolution, see \cite{dalsasso2020handle}).

\section{A multi-temporal extension of SAR2SAR}
\label{sec:mtsar2sar}
\subsection{Ratio-based filtering}
The proposed extension is based on RABASAR, a simple yet efficient multi-temporal despeckling method (see \cite{rabasar} for a complete description). The key idea is the use of the ratio between a speckled image and the temporal mean of the SAR image stack (named in the following "the \textit{super-image}" $s$). 

The first step of this method is to compute the super-image $s$. Averaging the multi-temporal series strongly reduces the speckle. If some noticeable speckle fluctuations remain, a slight spatial filtering may be necessary to obtain a very high-quality super-image. 

For any image $w$ in the stack, the ratio $\tau = {w}/{s}$ is then computed and denoised by a state-of-the-art speckle reduction method. In the absence of speckle fluctuations in the super-image, 
the noise distribution is similar to the distribution of a single-look SAR image and a traditional despeckling algorithm can be applied. If speckle fluctuations remain in the super-image image, an adaption of the despeckling algorithm is necessary (a generic method, called RuLoG, is proposed in \cite{rabasar}). For all despeckling methods that are sensitive to speckle correlations, a sub-sampling of the ratio image is necessary, which alters the resolution of the restored ratio image $\hat{\ratio}$.
Finally, the denoised estimation of $\reflectivity$ is retrieved by multiplying $\hat{\ratio}$ with the super-image $\superimage$: $\hat{\reflectivity} = \hat{\ratio} \times \superimage$. 

In this paper, we propose to replace the denoising step of the ratio image by the SAR2SAR approach, which avoids the sub-sampling procedure required by standard despeckling filters.



\begin{figure*}[p]
\centerline{\includegraphics[height=\textheight]{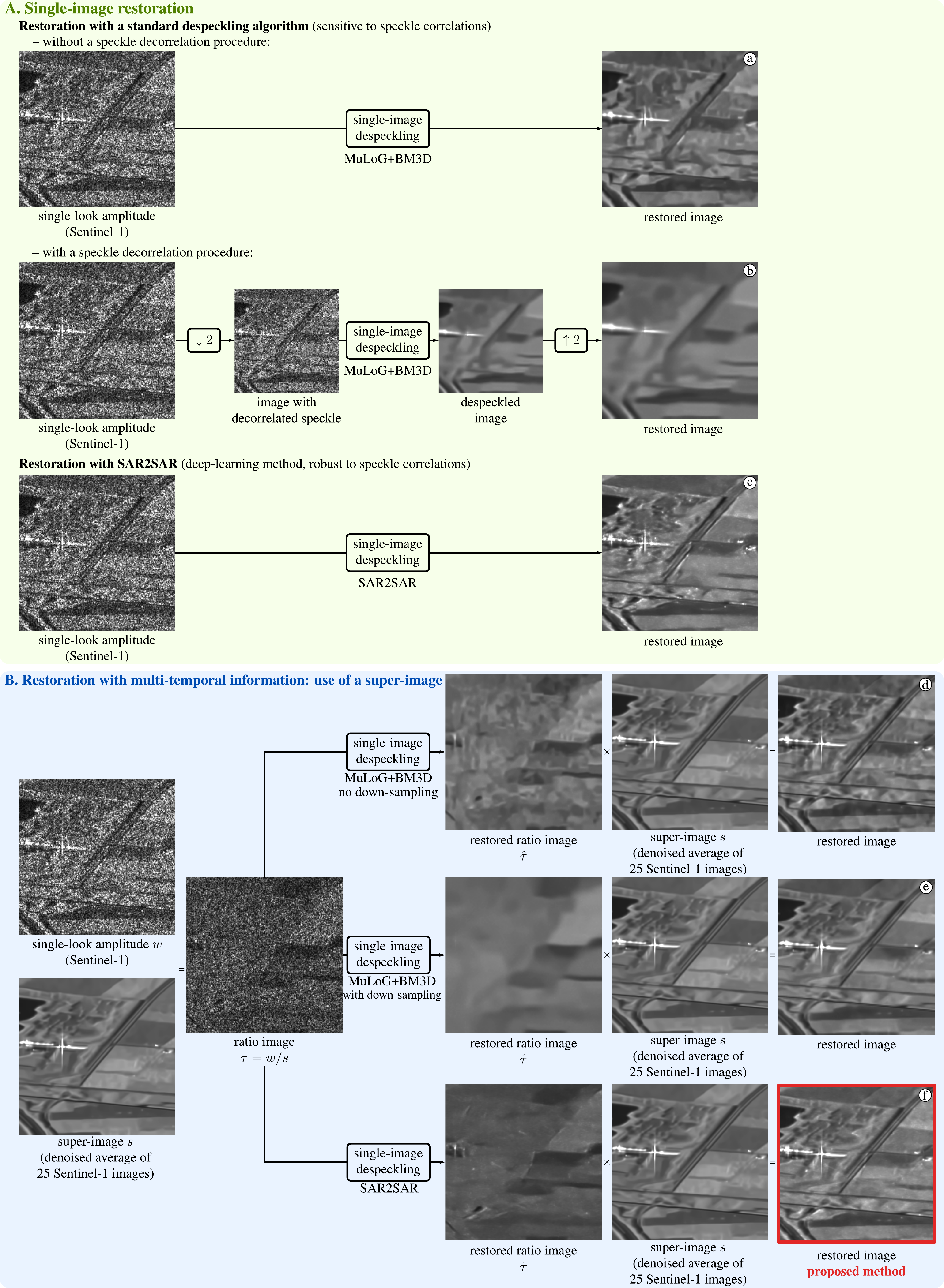}}\vspace*{-\baselineskip}
\caption{Comparison of several despeckling strategies for single-image and multi-temporal processing.}
\label{fig:allres}
\end{figure*}

\subsection{Adaptation of SAR2SAR to ratio images}
Owing to their non-linear nature, neural networks are 
very 
sensitive to the dynamic range of their inputs. Significantly shifting the 
dynamic range of input images between training and testing most often leads to 
catastrophic results. Ratio images have very different ranges compared to SAR 
intensity images: in the absence of significant changes between the image $w$ 
and the super-image $s$, the expected value of the ratio is 1. It is then 
necessary to appropriately rescale them in order to use the SAR2SAR network 
trained on SAR images (i.e., not specifically trained on ratio images). We 
experimented 
with several normalization strategies and describe the one that worked best. 
SAR2SAR processes log-transformed intensities $y=\log(w)$ that are 
approximately rescaled to the $[0,\,1]$ range by a fixed affine transform 
$y\mapsto (y-m)/(M-m)$ where $m$ and $M$ respectively correspond to the 
minimum and maximum log-transformed intensity computed over the whole training 
set. In order to preserve the original range, in log-domain, of the image $w$ 
when processing the ratio $\tau$, we normalize the super-image $s$: 
$s'=s/\lambda$. The normalization factor $\lambda$ is chosen such that the 
average log-transformed intensity of the super-image is equal to 0: 
$\lambda=\exp(\mu[\log(s)])$, with $\mu[\cdot]$ the average value computed 
over all pixels of an image. The \emph{modified} ratio image $\tau'=w/s'$ 
is 
processed by SAR2SAR. The obtained despeckled ratio $\hat{\tau}'$ is then 
multiplied by the normalized super-image to produce the final estimate of the 
restored image: $\widehat{w}=\hat{\tau}'\cdot s'$.

\section{Experimental results}
\label{sec:exp}


We illustrate our method on single-look Sentinel-1 images of an area near Lelystad, Netherlands. A stack of 25 images was spatially co-registered and temporally averaged. Remaining speckle fluctuations were suppressed with MuLoG+BM3D \cite{Dele-17}, using an equivalent number of look estimated in a homogeneous area. A single-look amplitude image and the super-image are shown in Fig.\ref{fig:allres}, left column.

Fig.\ref{fig:allres} compares restoration results obtained by several strategies: the top block gives single-image restoration results and the bottom block shows how the use of a super-image improves the despeckling. Images (a) and (d) suffer from artifacts due to the application of a despeckling method that is sensitive to spatial correlations of speckle directly on a Sentinel-1 image. Downsampling the images reduces speckle correlation and suppresses these artifacts ((b) and (e)). This comes at the cost of a noticeable resolution loss, somewhat mitigated by the use of a super-image. SAR2SAR is robust to speckle correlations. It gives superior results in the single-image scenario (image c) and offers a restoration with an improved preservation of details such as thin roads or field edges using the proposed multi-temporal approach (image f).

\section{Conclusion}
\label{sec:concl}
The combination of a deep neural network trained to remove speckle on actual SAR images and multi-temporal information in the form of a super-image (i.e., high-quality temporal average of co-registered SAR images) improves the quality of restored images. Future work will consider more complex ways to include multi-temporal information in deep-neural networks. Other methods to compute the super-image can also be considered, see \cite{Gasn-21}.

\small
\bibliographystyle{IEEEbib}
\bibliography{refs_register}

\begin{thebibliography}{10}

\bibitem{Good-76}
J.W. Goodman,
\newblock ``Some fundamental properties of speckle,''
\newblock {\em JOSA A}, 1976.

\bibitem{Dele-17}
C.~Deledalle, L.~Denis, S.~Tabti, and F.~Tupin,
\newblock ``{MuLoG, or How to apply gaussian denoisers to multi-channel SAR
  speckle reduction ?},''
\newblock {\em IEEE TIP}, 2017.

\bibitem{wang2017sar}
P.~Wang, H.~Zhang, and V.~M. Patel,
\newblock ``{SAR} image despeckling using a convolutional neural network,''
\newblock {\em IEEE SPL}, 2017.

\bibitem{wang2017generative}
P.~Wang, H.~Zhang, and V.~M. Patel,
\newblock ``Generative adversarial network-based restoration of speckled {SAR}
  images,''
\newblock in {\em IEEE CAMSAP}, 2017.

\bibitem{zhang2018learning}
Q.~Zhang, Q.~Yuan, J.~Li, Z.~Yang, and X.~Ma,
\newblock ``Learning a dilated residual network for {SAR} image despeckling,''
\newblock {\em Remote Sensing}, vol. 10, no. 2, pp. 196, 2018.

\bibitem{chierchia2017sar}
G.~Chierchia, D.~Cozzolino, G.~Poggi, and Luisa Verdoliva,
\newblock ``{SAR} image despeckling through convolutional neural networks,''
\newblock in {\em IGARSS}, 2017.

\bibitem{cozzolino2020nonlocal}
D.~Cozzolino, L.~Verdoliva, G.~Scarpa, and G.~Poggi,
\newblock ``Nonlocal {CNN} {SAR} {I}mage {D}especkling,''
\newblock {\em Remote Sensing}, vol. 12, no. 6, pp. 1006, 2020.

\bibitem{lattari2019deep}
F.~Lattari, B.~Gonzalez~Leon, F.~Asaro, A.~Rucci, C.~Prati, and M.~Matteucci,
\newblock ``Deep learning for {SAR} image despeckling,''
\newblock {\em Remote Sensing}, 2019.

\bibitem{dalsasso2020sar}
E.~Dalsasso, X.~Yang, L.~Denis, F.~Tupin, and W.~Yang,
\newblock ``{SAR} {I}mage {D}especkling by {D}eep {N}eural {N}etworks: from a
  pre-trained model to an end-to-end training strategy,''
\newblock {\em Remote Sensing}, vol. 12, no. 16, pp. 2636, 2020.

\bibitem{molini2020towards}
A.~B. Molini, D.~Valsesia, G.~Fracastoro, and E.~Magli,
\newblock ``Towards deep unsupervised {SAR} despeckling with blind-spot
  convolutional neural networks,''
\newblock {\em arXiv preprint arXiv:2001.05264}, 2020.

\bibitem{molini2020speckle2void}
A.~B. Molini, D.~Valsesia, G.~Fracastoro, and E.~Magli,
\newblock ``Speckle2{V}oid: {D}eep {S}elf-{S}upervised {SAR} {D}especkling with
  {B}lind-{S}pot {C}onvolutional {N}eural {N}etworks,''
\newblock {\em arXiv preprint arXiv:2007.02075}, 2020.

\bibitem{ma2020sar}
Xiaoshuang Ma, Chen Wang, Zhixiang Yin, and Penghai Wu,
\newblock ``{SAR} {I}mage {D}especkling by {N}oisy {R}eference-{B}ased {D}eep
  {L}earning {M}ethod,''
\newblock {\em IEEE TGRS}, 2020.

\bibitem{dalsasso2020sar2sar}
E.~Dalsasso, L.~Denis, and F.~Tupin,
\newblock ``{SAR2SAR}: a self-supervised despeckling algorithm for {SAR}
  images,''
\newblock {\em arXiv preprint arXiv:2006.15037}, 2020.

\bibitem{rabasar}
W.~Zhao, C.~Deledalle, L.~Denis, H.~Ma{\^i}tre, J.-M. Nicolas, and F.~Tupin,
\newblock ``{Ratio-Based Multitemporal SAR Images Denoising: RABASAR},''
\newblock {\em IEEE TGRS}, 2019.

\bibitem{dalsasso2020handle}
E.~Dalsasso, L.~Denis, and F.~Tupin,
\newblock ``{How to handle spatial correlations in {SAR} despeckling?
  Resampling strategies and deep learning approaches},''
\newblock {\em hal-02538046}, 2020.

\bibitem{Gasn-21}
N.~Gasnier, L.~Denis, and F.~Tupin,
\newblock ``{On the use and denoising of the temporal geometric mean for SAR
  time series},''
\newblock {\em {IEEE GRSL}}, 2021.

\end{thebibliography}

\end{document}